\documentclass[aps,prl,twocolumn,letterpaper,showpacs,10pt]{revtex4-1}
\usepackage[utf8]{inputenc}
\usepackage{graphicx,amsmath}
\usepackage{amssymb}
\usepackage{amsthm}

\begin{document}
\title{Charge conservation protected topological phases}
\author{Jan Carl Budich$^1$}

\affiliation{$^1$Department of Physics, Stockholm University, Se-106 91 Stockholm, Sweden;}
\date{\today}
\begin{abstract}
We discuss the relation between particle number conservation and topological phases. In four spatial dimensions, we find that systems belonging to different topological phases in the presence of a $\text{U}(1)$~charge conservation can be connected adiabatically, i.e., without closing the gap, upon intermediately breaking this local symmetry by a superconducting term. The time reversal preserving topological insulator states in 2D and 3D which can be obtained from the 4D parent state by dimensional reduction inherit this protection by charge conservation. Hence, all topological insulators can be adiabatically connected to a trivial insulating state without breaking time reversal symmetry, provided an intermediate superconducting term is allowed during the adiabatic deformation. Conversely, in one spatial dimension, non-symmetry-protected topological phases occur only in systems that break $\text{U}(1)$~charge conservation. These results can intuitively be understood by considering a natural embedding of the classifying spaces of charge conserving Hamiltonians into the corresponding Bogoliubov de Gennes classes.    
\end{abstract}
\maketitle

{\emph{Introduction}} --
In recent years, topological states of matter (TSM) that can be understood at the level of quadratic model Hamiltonians have become a major focus of condensed matter physics \cite{RyuLudwig,HasanKane,XLReview2010,TSMReview}. An exhaustive classification of all possible TSM in the ten Altland-Zirnbauer symmetry classes \cite{AltlandZirnbauer} of insulators and mean field superconductors has been achieved by different means in Refs. \cite{Schnyder2008,KitaevPeriodic,RyuLudwig}. For the symmetry class A of the quantum Hall effect in 2D, i.e., no symmetries except a local $\text{U}(1)$~charge conservation, there is a variety of topological phases apart from the integer quantum Hall (IQH) phases \cite{Klitzing1980,Laughlin1981,TKNN1982}, namely the family of fractional quantum Hall states \cite{StormerFQH,LaughlinState,PrangeGirvin} that exist only in the presence of interactions and hence cannot be adiabatically deformed into non-interacting band structures. These phases can be classified in the framework of topological order which was introduced by Wen back in 1990 \cite{WenTO}. In a more recent paper by Chen, Gu, and Wen \cite{WenLU}, it has been shown that different gapped phases which do not have local order parameters associated with spontaneous symmetry breaking must have different topological orders. Furthermore, Ref. \cite{WenLU} identifies different topological orders with different patterns of long range entanglement (LRE).\\

According to the definition of topological order given in Ref. \cite{WenLU}, non-interacting 2D band structures with different Chern numbers, i.e., various IQH states, have different topological orders since they concur in all symmetries and cannot be adiabatically connected \cite{Niu1985}. Clearly, the IQH states all have trivial topological entanglement entropy as defined in Refs. \cite{KitaevPreskill,LevinWenEntropy} as they all have quantum dimension one. Moreover, the wave functions of all IQH states are slater determinants which implies that the electrons are not entangled at all \cite{SchliemannLossEntanglement,Ghirardi2002,GhirardiPRA}. However, dividing such a many body system into two subsystems by virtue of a virtual cut in real space, a non-vanishing entropy in the reduced state of one subsystem might arise due to fluctuations in the particle number of this subsystem: If the wave function of one of the single particle states entering the slater determinant of occupied states is delocalized over the real space cut, it is plausible for the particle to be found in each of the subsystems. Hence, in this case, the reduced state of one subsystem will be a mixture of states with different particle number which gives the notion of a non-trivial entanglement entropy even for non-interacting states \cite{Peschel,TurnerTIEntanglement}-- a particle number entanglement. Along these lines a non-trivial Chern number is indeed in one-to-one correspondence to a long range particle number entanglement: A set of exponentially localized Wannier functions for a band can be found if and only if the Chern number of the band vanishes \cite{WannierChern}.\\

{\emph{Main results}}--
In this work, we view charge conservation as a protecting symmetry, i.e., we demonstrate that the statement whether two systems are adiabatically connected to each other can depend on the symmetry constraints related to particle number conservation. More specifically, we define the notion of a charge conservation protected topological phase (CPTP), i.e., a state that cannot be adiabatically connected to a trivial band structure in the presence of a locally conserved $\text{U}(1)$~charge but which can be connected to a trivial state without closing the gap if this $\text{U}(1)$~symmetry is intermediately broken down to a superconducting charge conservation modulo two. An example of a CPTP is the 4D analog of the quantum Hall effect \cite{ZhangHu4DQH} which is not symmetry protected in the conventional sense. The time reversal symmetry (TRS) preserving topological insulators (TI)s in 2D \cite{KaneMele2005a,KaneMele2005b,BHZ2006,Koenig2007} and 3D \cite{Fu3DTI,Hsieh2008,3DTIBiSe} are shown to be protected by both TRS and charge conservation. This means that a TI state can be adiabatically connected to a trivial state without breaking TRS if the constraint of $\text{U}(1)$~charge conservation is relaxed to a superconducting $\mathbb Z_2$~constraint. This should of course not be seen as a limitation of the robustness of these states since experimentally one will hardly pick up a superconducting term accidentally.\\   

{\emph{Embedding of charge conserving symmetry classes}} --
The symmetry class D can be formally obtained from the unitary class A by imposing a symmetry constraint, namely the anti-unitary particle hole symmetry (PHS) $\mathcal P$~with $\mathcal P^2=1$ (see Ref. \cite{Schnyder2008} and Tab. \ref{tab:classclean} which we repeat here for the reader's convenience). From this point of view, one might think of class D as a subset of class A. In 1D for example, there are no topological states in class A due to $\pi_1\left(G_{n,m}(\mathbb C)\right)=0$, where $G_{n,m}(\mathbb C)=U(n)/(U(m)\times U(n-m))$~denotes the complex Grassmannian. However, in the presence of PHS, topological band structures similar to the model introduced by Su, Schrieffer, and Heeger (SSH) \cite{SSH} can be defined that can only be adiabatically deformed into a trivial band structure if PHS is broken during the deformation. Therefore, the SSH model is called a symmetry protected topological state.\\

This picture needs revision if we think of a mean field superconductor without symmetries as a Bogoliubov-de Gennes (BdG) Hamiltonian in symmetry class D \cite{AltlandZirnbauer}. In this case, PHS is a consequence of the particle-hole redundancy present in the BdG picture rather than a physical symmetry: The BdG band structure consists of two copies of the underlying band structure where the energy spectrum of the hole bands is mirrored as compared to the equivalent electron bands. This enforces the presence of the spectrum generating PHS -- the hole bands are conjugated to the electron bands by PHS. This constraint is not a symmetry and cannot be broken physically. Hence the BdG analog of the SSH model should not be considered as a symmetry protected topological state since it cannot be connected to a trivial state without closing the gap.\\

\begin{figure}[bh]
\includegraphics[width=0.85\linewidth]{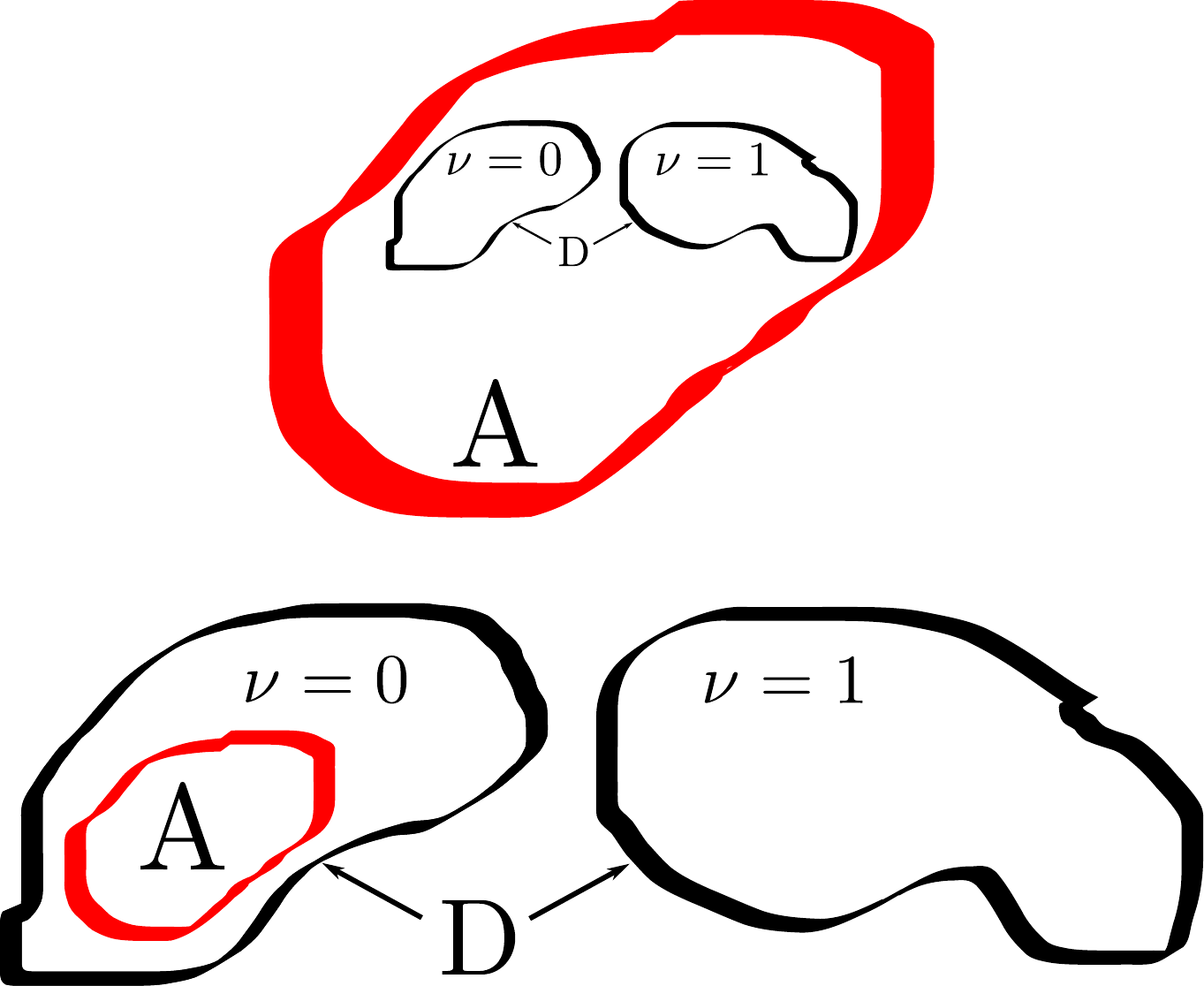}
\caption{(Color online)Top: Symmetry class D in 1D as a subset of charge conserving Hamiltonians in class A that preserve PHS with $\mathcal P^2=1$. The SSH model is in the connected component denoted by $\nu=1$. The superset A has only one connected component, i.e., all charge conserving states in 1D can be adiabatically connected to each other. Bottom: The symmetry class D as the set of BdG Hamiltonians without further symmetries in 1D has two connected components. The injective embedding $\mathcal E\left[ A \right]$~ of class A (red set denoted by A) defined in Eq. (\ref{eqn:embedding}) is a subset of the trivial connected component $\nu=0$. The BdG Hamiltonian of the Majorana chain is in the other connected component $\nu=1$.}
\label{fig:oned}
\end{figure}

Along these lines, symmetry class A can thus be viewed as a subset of all possible BdG Hamiltonians in class D: One can start with any non-interacting band structure in class A, create a hole-like copy of this band structure and is then even allowed to consider $\text{U}(1)$~charge conservation breaking superconducting terms which cannot be accounted for in symmetry class A. This viewpoint is equivalent to viewing class D as the set of arbitrary gapped Hamiltonians that are bilinear in Majorana fermion operators \cite{KitaevPeriodic}. Here, since we want to work in the BdG picture later on, 
we formalize this argument by an embedding map $\mathcal E: \text{A}\rightarrow \text{D}$~instead of going to the Majorana representation. This map is defined as
\begin{align}
H_0 \mapsto  \mathcal E(H_0)=
   \begin{pmatrix}
    H_0&0\\0&{-\mathcal T H_0\mathcal T^{-1}}
   \end{pmatrix},
   \label{eqn:embedding}
\end{align}
where $\mathcal T$~denotes the time reversal operation and $H_0 \in \text{A}$~is an arbitrary charge conserving gapped quadratic Hamiltonian. The resulting Hamiltonian $\mathcal E(H_0)$~ is a special case of a BdG Hamiltonian in class D that is still $\text{U}(1)$~charge conserving. This phenomenology is illustrated in Fig. \ref{fig:oned}. For later purposes, we note that if $H_0$~preserves TRS with $\mathcal T^2=-1$, i.e., if $H_0 \in \text{AII}$, then this restriction of the same map $\mathcal E$~defines an embedding of class AII into the BdG class DIII.\\

\begin{table}[ht]
\centering
\begin{tabular}{|l|l|cccc|}\hline
 Class &    symmetries    & $d=1$&$d=2$&$d=3$&$d=4$\\ \hline
 A & $(0,0,0)$&0&$\mathbb Z$&0&$\mathbb Z$\\
 D&$(0,+1,0)$&$\mathbb Z_2$&$\mathbb Z$&0&0\\
 DIII&$(-1,+1,1)$&$\mathbb Z_2$&$\mathbb Z_2$&$\mathbb Z$&0\\
 AII &$(-1,0,0)$&0&$\mathbb Z_2$&$\mathbb Z_2$&$\mathbb Z$\\ \hline
\end{tabular}
\caption{\label{tab:classclean} Table of topological invariants for the symmetry classes A, D, DIII, AII in spatial dimension $d=1\ldots 4$ ~\cite{RyuLudwig}. In the second column, the symmetries are denoted in the order $(\text{TRS},\text{PHS},\text{CS})$, where CS stands for chiral symmetry. The absence of a symmetry is denoted by $0$. For, TRS and PHS the entry $\pm 1$~denotes the presence of a symmetry operation which squares to $\pm 1$. The presence of CS is denoted by $1$.
}
\end{table}

A known consequence of this reasoning in 1D is the existence of non-symmetry-protected topological superconductors \cite{Kitaev2001} although there are no particle number conserving topological phases in the unitary class (see Fig. \ref{fig:oned} and Tab. \ref{tab:classclean}).
The BdG Hamiltonian of Kitaev's Majorana chain \cite{Kitaev2001} is similar to the SSH model. However, whereas a staggered potential breaks the particle hole symmetry of the SSH model, a formally equivalent term in the BdG Hamiltonian of the Majorana chain is forbidden by the fermionic algebra of the field operators. Hence, the SSH model is associated with the scenario illustrated in the upper panel of Fig. \ref{fig:oned}, whereas the set of Hamiltonians depicted in the lower panel of Fig. \ref{fig:oned} contains the Majorana chain. This example shows that relaxing the charge conservation symmetry from $\text{U}(1)$~to $\mathbb Z_2$~can give rise to new topological phases that are not symmetry protected.\\ 

Conversely, one might ask whether some topological phases are protected by the $\text{U}(1)$~charge conservation symmetry. This would be the case if a state could be adiabatically connected to a trivial insulating band structure in the presence of charge conservation but could be connected without closing the gap if we are allowed to break charge conservation by a superconducting term. In the following, we answer this question in the affirmative by discussing an example of a topologically non-trivial state in class A in 4D which we adiabatically connect to a trivial state by intermediately breaking charge conservation. We call such states charge conservation protected topological phases (CPTP). By dimensional reduction, we are able to show that also the 3D and 2D TI states are protected by charge conservation.\\       

\begin{figure}[bh]
\includegraphics[width=0.99\linewidth]{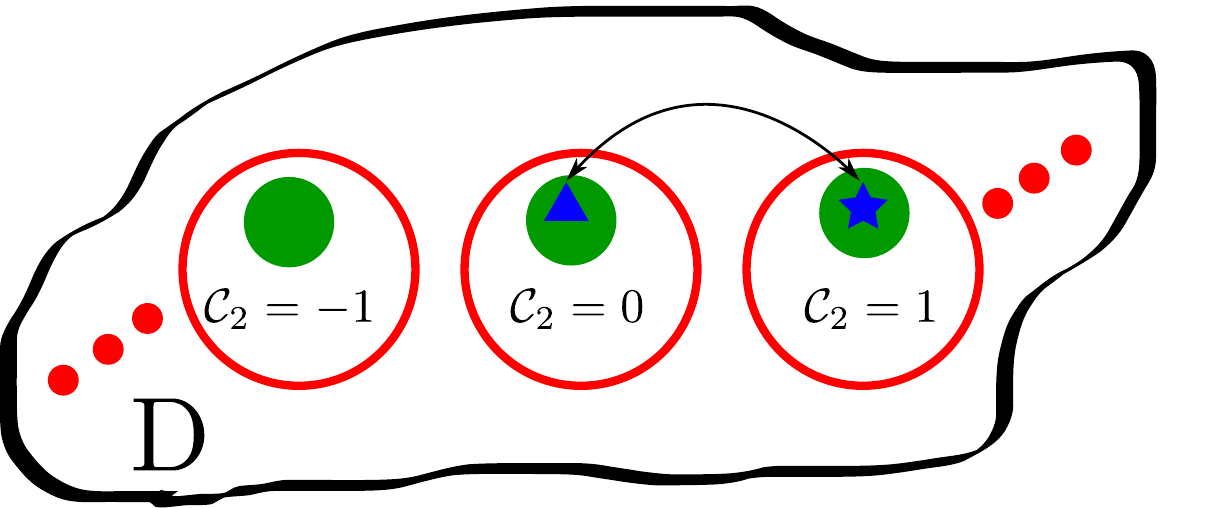}
\caption{(Color online) The set of BdG Hamiltonians without further symmetries in 4D has only one connected component. PHS conjugated copies of particle number conserving Hamiltonians with different second Chern number $\mathcal C_2$~are disconnected in symmetry class A but can be adiabatically connected within symmetry class D. Remarkably, for each value of $\mathcal C_2$, there is a non-empty subset of TRS preserving Hamiltonians (green). The blue star denotes the non-trivial topological state for $m<0$~whereas the blue triangle denotes the trivial state for $m>0$. The arrow between the two states sketches our gapped interpolation via an intermediate superconducting state.}
\label{fig:fourd}
\end{figure}

{\emph{A U(1) protected phase in 4D}}--
In 4D, symmetry class A is characterized by an integer topological invariant, the second Chern number $\mathcal C_2$~which distinguishes different topological phases. However, class D is trivial in 4D, i.e., all BdG Hamiltonians are equivalent (see Fig. \ref{fig:fourd} and Tab. \ref{tab:classclean}).
We hence suspect that different topological phases can be connected adiabatically by intermediately breaking the $\text{U}(1)$~charge conservation down to $\mathbb Z_2$. We now explicitly construct such an adiabatic interpolation. Let us start from a toy model for the 4D analog of the quantum Hall effect \cite{ZhangHu4DQH,QiTFT}. The model Hamiltonian reads
\begin{align}
&H_0=d^\mu\Gamma_\mu\nonumber\\
&d^0 = m+ k^2,~d^i=k_i,~i=1,\ldots,4
\label{eqn:4DAHam}
\end{align}
where the 4$\times$4 Dirac matrices $\Gamma_\mu$~are given by
\begin{align}
 &\Gamma_0=s_0\otimes \sigma_z,~\Gamma_1=s_0\otimes\sigma_y,~\Gamma_2=s_x\otimes\sigma_x\nonumber\\
 &\Gamma_3=s_y\otimes\sigma_x,~\Gamma_4=s_z\otimes\sigma_x
\end{align}
and satisfy the $SO(5)$~Clifford algebra \cite{AvronQuadrupole1989,Demler1999}
\begin{align}
\left\{\Gamma_\mu,\Gamma_\nu\right\}=2\delta_{\mu\nu}.
\label{eqn:ciff5}
\end{align}
For large momenta, the isotropic $k^2$~term dominates the Hamiltonian. We can thus compactify the $k$-space of our model to $S^4$~by identifying $k\rightarrow \infty$~with a single point. The second Chern number
\begin{align}
\mathcal C_2 =-\frac{1}{8\pi^2}\int_{S^4}\text{Tr}\left[\mathcal F\wedge \mathcal F\right]
\end{align}
with the non-Abelian Berry curvature $\mathcal F$~is then integer quantized, and it has been shown \cite{QiTFT} that $\mathcal C_2=1$~for $m<0$~whereas $\mathcal C_2 =0$~for $m>0$. These two phases cannot be connected to each other without closing the bulk gap as long as particle number conservation is preserved. However, by adiabatically switching on a superconducting gap, we will be able to connect the two states without ever closing the bulk gap. We first introduce a particle hole pseudo spin $\tau$. The BdG Hamiltonian associated with our model then reads
\begin{align}
H=\begin{pmatrix}
    H_0&0\\0&{-\mathcal T H_0\mathcal T^{-1}}
   \end{pmatrix}=H_0 \tau_z,
\end{align}
where we have used the time reversal invariance of $H_0$~under $\mathcal T=is_y K$~in the last step. In this basis, the emergent PHS operation takes the form $\mathcal P=s_y \tau_y K$, where $K$~denotes complex conjugation. We now switch on the fictitious superconducting term $\Delta(\lambda)\tau_x$, where $\lambda \in\left[0,\pi\right]$~is the parameter of the adiabatic interpolation and $\Delta(\lambda)=\sin(\lambda)$. Furthermore, we make $H_0$~dependent on $\lambda$~by defining $d^0(\lambda)=\cos(\lambda)+k^2$, i.e., the Dirac mass parameter $m$~acquires the $\lambda$-dependence $m(\lambda)=\cos(\lambda)$. Obviously, $H_0(0)$~is the trivial insulator with $\mathcal C_2=0$~whereas $H_0(\pi)$~is the non-trivial insulator with $\mathcal C_2=1$. The additional superconducting term vanishes at $\lambda=0,\pi$. Hence, both starting and end point of the interpolation preserve $\text{U}(1)$~charge conservation. The spectrum of the total BdG Hamiltonian $\tilde H(\lambda)=H(\lambda)+\Delta(\lambda)\tau_x$~can be conveniently obtained by taking the square:
\begin{align}
E^2&= \tilde H(\lambda)^2=\lvert d\rvert^2+\left\{H_0\tau_z,\Delta\tau_x\right\}+\Delta^2=\nonumber\\
&=\lvert d\rvert^2+\Delta^2\ge 1~ \forall \lambda,k.
\end{align}
This interpolation is fully gapped and describes a formally well defined BdG Hamiltonian for all values of the interpolation parameter $\lambda$~since $\tau_x$~preserves the emergent PHS operation in our choice of basis. Thus we have connected two different topological phases adiabatically with the help of an intermediate superconducting term. \\

{\emph{Topological insulators as CPTP}}--
In Eq. (\ref{eqn:4DAHam}), we could without loss of generality choose a model Hamiltonian that preserves TRS and is hence not only in symmetry class A but also in AII, the symplectic class. This is because in 4D a Hamiltonian with an arbitrary second Chern number can be adiabatically deformed into a representative that preserves TRS  within in the same topological equivalence class, i.e., without changing the second Chern number (see Fig. \ref{fig:fourd}). This makes the 4D analog of the IQH effect \cite{ZhangHu4DQH} the parent state of a dimensional hierarchy of topological states \cite{QiTFT,RyuLudwig,XLReview2010,TSMReview}. The two lower dimensional descendants in class AII are the 3D TI \cite{Fu3DTI,Hsieh2008}, and the 2D TI a.k.a. the quantum spin Hall state \cite{KaneMele2005a,KaneMele2005b,BHZ2006,Koenig2007}. In contrast to the parent state, these states are protected by TRS, i.e., they can be adiabatically connected to a trivial state if TRS is broken. We will now show that in addition they are protected by charge conservation in the same sense as their parent state. To this end, we express the model for the 3D TI presented in Ref. \cite{3DTIBiSe} and the model for the quantum spin Hall effect introduced in Ref. \cite{BHZ2006} as dimensional reductions from the toy model for the 4D parent state (\ref{eqn:4DAHam}). Explicitly, by setting $d^1=0$~in Eq. (\ref{eqn:4DAHam}), we obtain a toy model similar to those in Refs. \cite{3DTIBiSe,XLReview2010} for the 3D TI which is non-trivial for $m<0$,~and by setting $d^2=d^3=0$~we obtain a minimal model for the quantum spin Hall effect which is very similar to the one presented in Ref. \cite{BHZ2006}. It then follows that for these dimensional reductions, the gapped interpolation with the same superconducting term $\Delta\tau_x$~as shown above for the parent state can be performed in complete analogy. This concludes our proof that both the 2D and the 3D TI state are protected by $\text{U}(1)$~charge conservation. Note that this conclusion cannot be obtained from looking at the periodic table of topological states \cite{KitaevPeriodic,RyuLudwig} (see Tab. \ref{tab:classclean}). This is due to the existence of different topological phases also in class DIII which are characterized by a $\mathbb Z_2$~invariant in 2D and a $\mathbb Z$~invariant in 3D, respectively \cite{Schnyder2008,RyuLudwig}. Whereas, as we have just shown, the TI states belong to the trivial phase in DIII, there are TRS preserving topological superconductors \cite{RoyTSC,Schnyder2008,QiTSC} which represent the non-trivial phases in DIII.\\

{\emph{Conclusions}}--
We embedded the particle number conserving symmetry classes A and AII into the corresponding BdG classes D and DIII, respectively. In 4D, states that are in different topological phases as long as the particle number is conserved can be adiabatically connected in the BdG class D. By dimensional reduction from the 4D parent state, we could show that topological insulators in the symplectic class AII in 2D and 3D can be adiabatically connected to trivial insulating states without breaking TRS if the $\text{U}(1)$~charge conservation is relaxed to the superconducting charge conservation modulo two. In 2D, BdG Hamiltonians consisting of PHS conjugated copies of different IQH states have different Chern numbers as well. Therefore, different IQH states in 2D cannot be adiabatically connected to each other with the help of a superconducting term. Our analysis shows that charge conservation can play the role of a protecting symmetry for topological phases.\\

{\emph{Acknowledgments}}--
We would like to thank Eddy Ardonne, Hans Hansson, Patrik Recher, and Bjoern Trauzettel for fruitful discussions as well as the Swedish Science Research Council for financial support.

\end{document}